\documentclass[aps,superscriptaddress,floats,showpacs,floatfix,twocolumn]{revtex4}
\usepackage{bm}
\usepackage{graphicx}
\usepackage{subfigure}
\usepackage{amssymb}
\usepackage{amsmath}
\usepackage{color}
\usepackage[a4paper,left=1.5cm, right=1.3cm, top=2cm,bottom=2cm]{geometry}
\begin{document}
\newcommand{\joerg}[1]{\textcolor{red}{#1}}
\newcommand{\janet}[1]{\textcolor{blue}{#1}}

\title{Turbulent superstructures in Rayleigh-B\'{e}nard convection}
\author{Ambrish Pandey}
\affiliation{Institut f\"ur Thermo- und Fluiddynamik, Technische Universit\"at Ilmenau, Postfach 100565, D-98684 Ilmenau, Germany}
\author{Janet D. Scheel}
\affiliation{Department of Physics, Occidental College, 1600 Campus Road, M21, Los Angeles, California 90041, USA}
\author{J\"org Schumacher}
\affiliation{Institut f\"ur Thermo- und Fluiddynamik, Technische Universit\"at Ilmenau, Postfach 100565, D-98684 Ilmenau, Germany}
\date{\today}

\begin{abstract}
Turbulent Rayleigh-B\'{e}nard convection displays a large-scale order in the form of rolls and cells on lengths larger than the layer height 
once the fluctuations of temperature and velocity are removed. These turbulent superstructures are reminiscent of the patterns 
close to the onset of convection. They are analyzed by numerical simulations of turbulent convection in fluids at different Prandtl 
number ranging from 0.005 to 70 and for Rayleigh numbers up to $10^7$.  For each case, we identify characteristic scales 
and times that separate the fast, small-scale turbulent fluctuations from the gradually changing large-scale superstructures. 
The characteristic scales of the large-scale patterns, which change with Prandtl and Rayleigh number, are also found to be correlated 
with the  boundary layer dynamics, and in particular the clustering of thermal plumes at the top and bottom plates. Our analysis suggests 
a scale separation and thus the existence of a simplified description of the turbulent superstructures in geo- and astrophysical settings.  
\end{abstract}
\maketitle

Large temperature differences across a horizontally extended fluid layer induce a turbulent convective fluid motion which is relevant in 
numerous geo- and astrophysical systems \cite{Kadanoff2001}.  These flows are typically highly turbulent with very large Rayleigh 
numbers $Ra$, the parameter that quantifies the intensity of the thermal driving in convection. From the classical 
perspective of turbulence one would expect a chaotic, irregular motion of differently sized vortices and thermal plumes. 
Rather than such a featureless stochastic fluid motion, some turbulent flows in nature display an organization into prominent and 
regular flow patterns that persist for times 
long compared to an eddy turnover time and extend over lengths which are larger than the height scale.
Examples are cloud streets in the atmosphere \cite{Markson1975} or granulation 
networks at the solar surface \cite{Nordlund2009} and other stars \cite{Michel2008}. This large-scale order will be termed a turbulent 
superstructure. It is observed in turbulent convection flows with very different molecular dissipation properties. The Prandtl 
number $Pr=\nu/\kappa$, another dimensionless parameter which relates kinematic viscosity $\nu$ to temperature diffusivity $\kappa$, 
is for example very small for stellar convection, $Pr\lesssim 10^{-3}$ \cite{Spiegel1962,Thual1992,Hanasoge2016}. It is 0.7 for atmospheric 
flows  and 7.0 for heat transport in the oceans. Rayleigh-B\'{e}nard convection (RBC) is the simplest turbulent convection flow 
evolving in a planar fluid layer of height $H$ that is uniformly heated with a temperature $T=T_b$ from below and cooled from 
above with $T=T_t$ such that $T_b-T_t=\Delta T>0$. The Rayleigh number is given by $Ra=g\alpha \Delta T H^3/(\nu\kappa)$ with 
$g$ being the acceleration due to gravity and $\alpha$ the thermal expansion coefficient. RBC can be considered as a paradigm for 
many applications \cite{Ahlers2009,Chilla2012} that usually contain further physical processes, such as radiation \cite{Christensen1996} 
and phase changes \cite{Stevens2005,Pauluis2011}, and additional fields such as magnetic fields \cite{Aurnou2010}.   
Numerical simulations of convection \cite{Hartlep2003,Hartlep2005,Rincon2005,Hardenberg2008,Bailon2010,Emran2015} have enabled 
researchers to access the large-scale structure formation in turbulent convection flows. Long-term investigations at very small 
Prandtl numbers $Pr\ll 0.1$ require simulations on massively parallel supercomputers in order to resolve the highly inertial turbulence properly. 
Such simulations have not been done before and this is a central motivation for the present study.   

At the onset of convection, $Ra_c=1708$, straight convection rolls have a unique and Prandtl-number-independent wavelength, 
$\lambda_c\approx 2H$ \cite{Jeffreys1928,Chandrasekhar1961}. For $Ra\gtrsim Ra_c$, these rolls become susceptible to secondary 
linear instabilities causing modulations, such as Eckhaus, zig-zag or oscillatory patterns \cite{Busse1978,Cross1993,Bodenschatz2000}. 
These secondary instabilities depend strongly on the Prandtl number of the working fluid and the wavenumber range of the plane-wave 
perturbation to the convection straight rolls in the layer \cite{Busse1978}. Dependencies on Rayleigh and Prandtl numbers of the pattern 
wavelength for $Ra>Ra_c$ have been studied systematically in RBC experiments in air, water and silicone oil by Willis et al. \cite{Willis1972}. 
Average roll widths tend to increase with $Ra$, which the authors attributed to increasingly unsteady three-dimensional motions. The trend with growing 
$Pr$ is less systematic \cite{Hartlep2003} and accompanied by hystereses at $Pr\gg 1$ \cite{Willis1972}. 

Roll and cell patterns of the velocity field in a {\em turbulent} RBC for $Ra\gtrsim 10^5$ that are reminiscent of the flow structures in the weakly 
nonlinear regime at $Ra \lesssim 5\times 10^3$ have been observed in recent DNS at $Pr\gtrsim 1$ \cite{Bailon2010,Emran2015}. Their 
detection requires an averaging over a time interval that should be long enough to remove the turbulent fluctuations in the 
fields effectively and yet short enough to not wash away the large-scale structures \cite{Emran2015}. A sliding time average
with an appropriate time window width should thus be able to separate the fast, small-scale turbulent fluctuations of velocity and temperature 
from the gradual variation of the large-scale superstructure patterns. Physically, this time window should be connected with the turnover time of fluid 
parcels in the superstructure rolls and cells. The determination of this averaging time scale as a function of $Ra$ and $Pr$ is a second motivation 
for the present study.
\begin{figure}
\centering
\includegraphics[scale = 0.4]{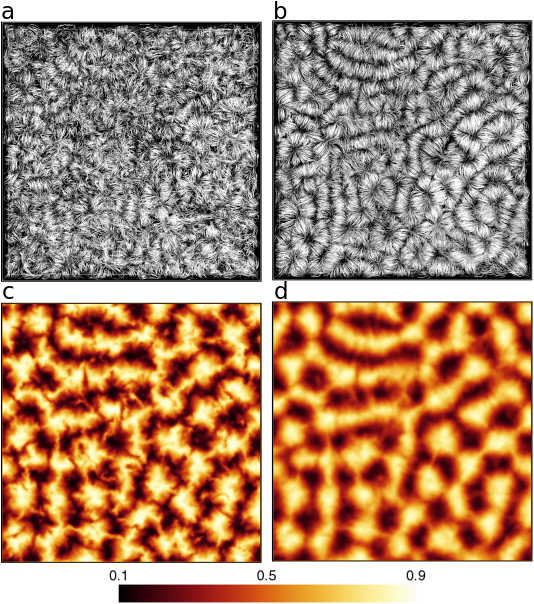}
\caption{{\bf Instantaneous and time-averaged fields.} Field line plots of the instantaneous (a) and time-averaged (b) velocity for
one of the lowest $Pr$ in our simulations. View is from the bottom of the layer. Corresponding instantaneous (c) and time-averaged (d)
temperature field. Data are for a Rayleigh number $Ra=10^5$ and $Pr=0.021$. Averaging time is 27 free-fall times $T_f$. The three-dimensional simulation 
domain is resolved by more than 1.2 billion mesh cells.}
\label{fig0}
\end{figure}
\begin{figure*}
\begin{center}
\includegraphics[scale = 0.4]{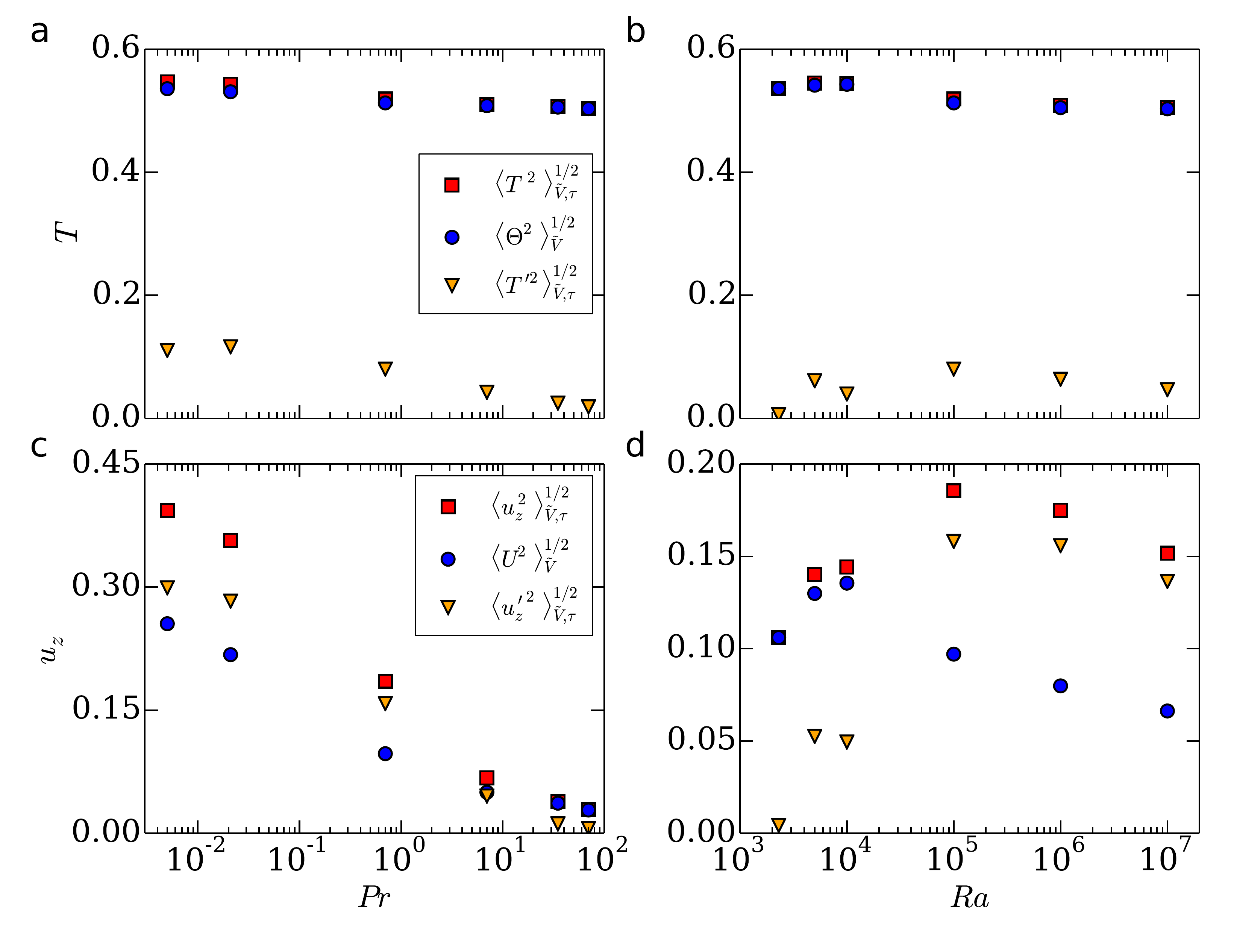}
\caption{{\bf Magnitude of turbulent superstructures.} The panels compare the root mean square (rms) values of the full field, 
$\langle T^2\rangle_{\tilde V,\tau}^{1/2}$ and $\langle u_z^2\rangle_{\tilde V,\tau}^{1/2}$, the temporal  means 
$\langle \Theta^2\rangle_{\tilde V}^{1/2}$ and $\langle U^2\rangle_{\tilde V}^{1/2}$ as defined in Eqns. (\ref{uvf},\ref{tvf}), 
and the fluctuations about the temporal mean, $\langle {T^{\prime}}^2\rangle_{\tilde V,\tau}^{1/2}$ and $\langle {u^{\prime}_z}^2\rangle_{\tilde V,\tau}^{1/2}$, 
for the temperature $T$ in panels (a, b) and for the vertical velocity component $u_z$ in panels (c, d). 
The dependence on $Pr$ at $Ra=10^5$ is given in panels (a, c) and on $Ra$ at  $Pr=0.7$ in panels (b, d). The data for the temporal means 
are shown in Fig. \ref{fig1}.}
\label{fig2}
\end{center}
\end{figure*}

In the present work, we report an analysis of the characteristic spatial and temporal scales of turbulent superstructures in RBC 
by means of three-dimensional direct numerical simulations (DNS) spanning more than four orders of magnitude in $Pr$ 
and more than three orders in $Ra$. All simulations reported here are of the Boussinesq equations of motion and performed 
in an extended closed square cell of aspect ratio of 25:25:1.  We identify the characteristic averaging time scales, $\tau(Ra, Pr)$, which will be
connected with a characteristic spatial scale (or wavelength) that can be determined by a spectral analysis of the turbulent 
superstructures. Our study of large-aspect-ratio turbulent RBC extends to very small Prandtl numbers with values 
significantly below 0.1, which have not been obtained before. The gradual evolution of the patterns at all Prandtl numbers is confirmed 
by radially averaged, azimuthal power spectra that reveal a gradual switching of the orientation of the superstructures which is 
reminiscent of cross-roll or skewed varicose instabilities that are well-known from the weakly nonlinear regime of RBC. Furthermore,
we compare the characteristic pattern scale in the bulk of the RBC flow to the scales of plumes and plume clusters that are present in the boundary layers in the vicinity of the top and bottom walls. The temperature patterns in the bulk are found to be correlated with the most 
prominent ridges in the vertical temperature field derivative at the bottom and top plates which in turn are correlated with the wall stresses
of the advecting velocity. Our analysis provides characteristic separation time and length scales for turbulent convection flows in extended domains
and thus opens the possibility to describe the superstructure patterns in turbulent convection by effective and reduced models that separate the fast, 
small scales from the slow, large scales. These reduced models can advance our understanding of a variety of turbulent systems that exhibit large-scale pattern formation, including mesoscale convection and solar granulation. 

\section*{Results}

\noindent
{\bf Superstructures for different Rayleigh and Prandtl numbers.}
Figure \ref{fig0} shows the velocity field lines (top row) and the corresponding temperature contours in the midplane (bottom row) for 
a simulation at one of the lowest Prandtl numbers in our simulations. While the instantaneous pictures display the expected irregularity of a turbulent flow as 
visible for example by the streamline tangle in panel (a), the averaged data reveal a much more ordered pattern. We also see that 
the superstructure patterns are more easily discerned in 
temperature field snapshots than in those of the velocity field. Figure \ref{fig2} confirms this observation. Here, we plot the root mean square (rms) 
values of the vertical velocity component $u_z$ and the temperature $T$. In agreement with Fig. \ref{fig0}, we split both fields into contributions 
coming from the time average over the time interval $\tau$ and the fluctuations,
\begin{align}
u_z({\bm x},t)&=U({\bm x})+u_z^{\prime}({\bm x},t)\,,\\ 
T({\bm x},t)&=\Theta({\bm x})+T^{\prime}({\bm x},t)\,.
\end{align}
The averaging volume $\tilde{V}$ is a slab around the midplane. See Eqns. (\ref{uvf}) and (\ref{tvf}) later in the text for definitions of $U$ and $\Theta$.
It can be seen that the rms values of the total and time averaged temperature are always
close together when Prandtl and Rayleigh number are varied. This is in contrast to the vertical velocity component. Fluctuations dominate here when the Prandtl 
numbers are low and the Rayleigh numbers are sufficiently high. An averaging with respect to time is thus necessary to reveal the patterns for 
both turbulent fields.        
  
\begin{figure*}
\centering
\includegraphics[width=0.9\textwidth]{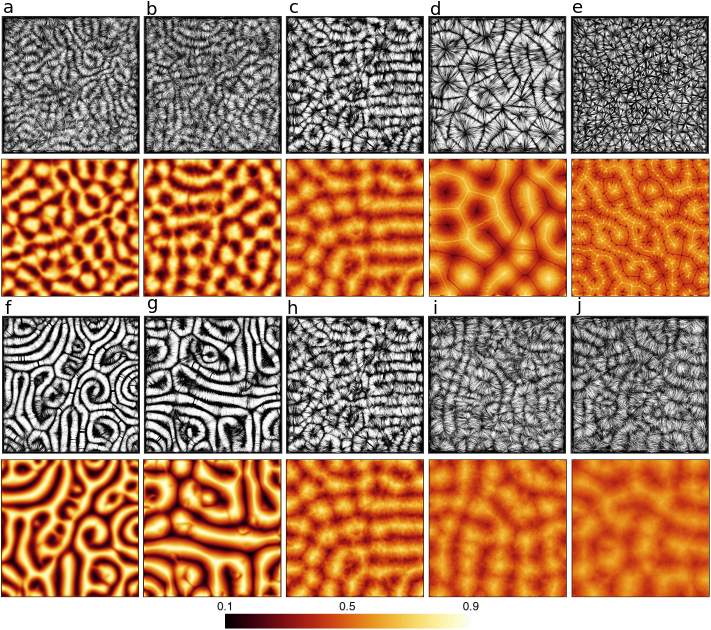}
\caption{{\bf Turbulent superstructures at different Rayleigh and Prandtl numbers.} For each of the simulations field line plots of the 
time-averaged velocity (top rows) and the corresponding time-averaged temperature in the midplane (bottom rows) are displayed. (a) turbulent Rayleigh-B\'{e}nard 
convection at a Rayleigh number $Ra=10^5$ at  $Pr=0.005$, (b) at $Pr=0.021$, (c) at $Pr=0.7$, (d) at $Pr=7$, and (e) at $Pr=70$. 
Averaging times in (a-e) are 21 free-fall times $T_f$ for $Pr=0.005$, $27 T_f$ for $Pr=0.021$, $57 T_f$ for $Pr=0.7$, $207 T_f$ for $Pr=7$, 
and $375 T_f$ for $Pr=70$. (f) turbulent Rayleigh-B\'{e}nard convection at a Prandtl number of $Pr=0.7$ at $Ra=5 \times 10^3$, (g) at $Ra=10^4$, (h) at $Ra=10^5$ 
(same panels as in (c)), (i) at $Ra=10^6$, and (j) at $Ra=10^7$. The averaging times are now $ 54 T_f$ for $Ra=5\times 10^3$, $72 T_f$ for $Ra=10^4$, $57 T_f$ for 
$Ra=10^5$, $66 T_f$ for $Ra=10^6$, and $72 T_f$ for $Ra=10^7$. All shown cross sections are $25 H\times 25H$ with $H$ being the 
height of the convection layer.}
\label{fig1}
\end{figure*}

Figure \ref{fig1} displays velocity field lines and temperature contours of time-averaged turbulent RBC flows at Prandtl number ranging from 
$Pr=0.005$ to 70 at $Ra=10^5$ and at Rayleigh number ranging from $Ra=5\times 10^3$ to $10^7$ for convection in air at $Pr=0.7$. All runs are 
turbulent and thus beyond the weakly nonlinear regime, except the runs in panel (e) at $Pr=70$, panel (f) at $Ra=5000$, and panel (g) at $Ra=10^4$
respectively.
For the non-turbulent cases the time averaged data does not deviate significantly from the instantaneous snapshots. If we look at the trends for all 
runs, we see that the velocity field lines form curved rolls for the lower $Pr$ and cell-like patterns for $Pr\ge 7$.  These structures fill the whole layer 
and are reminiscent of patterns at the onset of convection at much smaller Rayleigh numbers \cite{Bodenschatz2000}.
The corresponding temperature averages in the midplane show alternating ridges of cold downwelling and hot upwelling fluid which are 
coarser for the lowest Prandtl numbers and the highest Rayleigh numbers, respectively. For $Pr=0.005$ and 0.021, this is due to the highly diffusive 
temperature field that is in conjunction with an inertia-dominated fluid turbulence \cite{Schumacher2015,Schumacher2016,Scheel2016}. In case 
of the highest Prandtl number, $Pr=70$ at $Ra=10^5$, the amplitude of the turbulent velocity field fluctuations is significantly smaller and the 
temperature field displays much finer filaments. Coarser temperature patterns can also be observed for the highest Rayleigh number at $Ra=10^7$.
In the Supplementary Material, we plot additional vertical profiles of the velocity fluctuations as well as list further details for all simulation runs and in the  
Methods section the characteristic units are given which we use to formulate the Boussinesq model in dimensionless form. While 
low-Prandtl-number convection transports momentum very efficiently, the heat transport becomes significantly larger at the higher Prandtl numbers. 
Figure \ref{fig1} also demonstrates that the characteristic mean width of the rolls and spirals varies with $Pr$ and $Ra$. 

\vspace{0.5cm}
\noindent
{\bf Characteristic times and scales of superstructures.}
The free-fall time $T_f=(H/g\alpha\Delta T)^{1/2}$ is a characteristic convective time unit that stands for the (relatively) fast dynamics of thermal 
plumes and larger vortices in a turbulent convection flow. A slower time unit in the turbulent flow is either a vertical viscous ($Pr<1$) or a vertical 
diffusive ($Pr>1$) time composing an effective dissipative time by $T_d=\max(t_{\kappa}, t_{\nu})$ with $t_{\kappa}=H^2/\kappa$ and $t_{\nu}=H^2/\nu$. 
A complete removal of the large-scale patterns would require an averaging period on the order of $\Gamma^2T_d$ (with $\Gamma$ being the 
aspect ratio of the domain) which is $\gg 10^3 -10^4 T_f$, i.e., times which are not accessible in our massively parallel turbulence simulations.

\begin{figure*}
\begin{center}
\includegraphics[scale = 0.32]{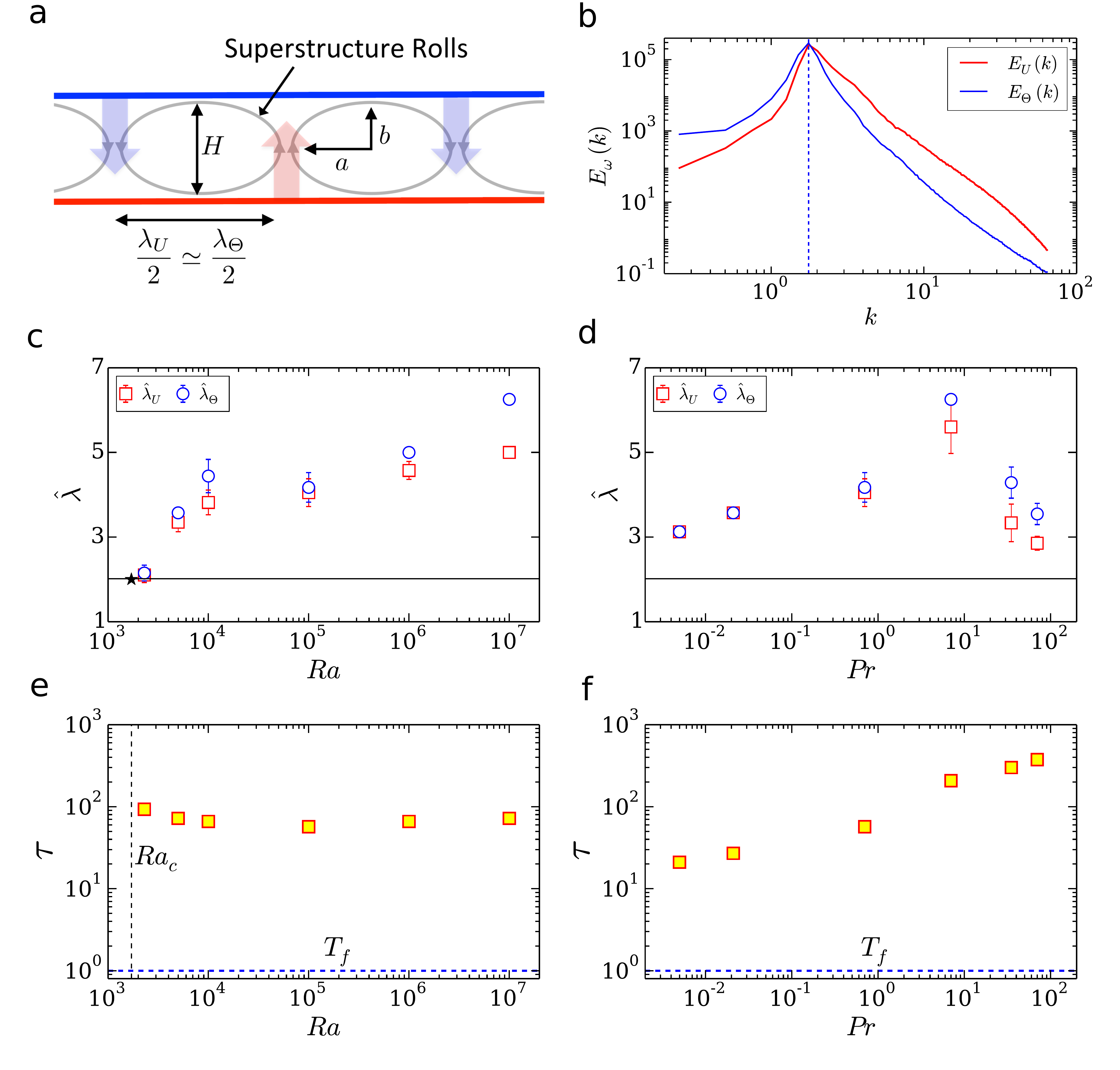}
\caption{{\bf Characteristic times and scales of turbulent superstructure patterns.} (a) Sketch of the turbulent superstructure with the rolls (or cells) 
formed by upwelling hot and downwelling cold fluid. The charactersitic scale of the superstructures is half the roll pattern wavelength, $\hat{\lambda}_{U,\Theta}$. 
(b) Azimuthally averaged power spectra, $E_{U}(k)$ and $E_{\Theta}(k)$. Data are for $Pr=0.021$ and $Ra=10^5$ as an example. The power spectra are 
determined for each snapshot and then averaged over all snapshots.
Maximum wavenumbers of both power spectra are indicated by the vertical dotted 
line. (c)  Rayleigh number dependence of characteristic wavelength $\hat{\lambda}_{U,\Theta}$ of the superstructures for turbulent convection in air. (d) 
Prandtl number dependence of characteristic wavelength of superstructures at $Ra=10^5$. All error bars are plotted though some are too small to see. 
The error bars are given by $\pm \Delta \lambda_{U,\Theta}= (2\pi/k^{\ast\,2}_{U,\Theta}) \Delta k$ with $\Delta k=\max(k^{\ast}_{U,\Theta}(t_0))-
\min(k^{\ast}_{U,\Theta}(t_0))$. The solid lines in panels (c, d) mark the critical wavelength $\lambda_c=2\pi/k_c$ with $k_c=3.117$ 
at the onset of convection \cite{Chandrasekhar1961}. (e) Characteristic time $\tau$ as a function of Rayleigh number at $Pr=0.7$. (f) Characteristic 
time as a function of the Prandtl number at $Ra=10^5$. The free-fall time $T_f$ is also indicated in both panels as a horizontal dashed line. The vertical dashed line in panel (e) stands for $Ra=Ra_c=1708$.}
\label{fig3}
\end{center}
\end{figure*}

Thus, the  averaging time $\tau$ that separates small-scale turbulence and superstrucutres should be bounded by
\begin{equation}
T_f \ll \tau(Ra, Pr) \ll T_d\,.
\end{equation}
This time $\tau$ should be considered as a representative value of a finite range of times rather than an exact time and is expected to show a 
dependence on our two system parameters $Ra$ and $Pr$. In the Supplementary Material it is shown for two different Prandtl numbers how the 
patterns change when the averaging time is varied. On the one hand, $\tau$ should be long enough to remove all small-scale fluctuations and to 
reveal the superstructures, in particular of velocity. On the other hand, $\tau$ has to be short enough such that the large-scale patterns are not 
removed completely.  Hence we define $\tau$ as the characteristic turnover time of  fluid parcels in the circulation rolls or cells, the latter of which 
extend across the whole layer from bottom to top and are considered as the building blocks of the superstructure velocity patterns.

In order to proceed, we decompose the RBC fields into a fast changing and gradually evolving contribution. This is inspired by asymptotic 
expansions that are developed for constrained turbulence, e.g., fast rotation or strong magnetic fields \cite{Julien2007,Klein2010,Malecha2014}. 
Furthermore, we substitute the full temperature field, $T({\bm x},t)$, by its deviation from the linear diffusive equilibrium profile, 
$\theta({\bm x},t)=T({\bm x},t)-T_{\text{lin}}(z)$. Our focus is on the horizontal patterns in the system. Therefore, the 
subsequent superstructure analysis is focussed on the symmetry plane at $z=1/2$ where the patterns are identified by upwelling 
hot and downwelling cold fluid (see Fig. \ref{fig3}(a)). The  gradually varying fields are given by the following sliding time average 
with respect to $\tau$ 
\begin{align}
    \label{uvf}
    U(x,y;\tau,t_0) &=\frac{1}{\tau}\int^{t_0+\tau/2}_{t_0-\tau/2} u_z(x,y,z=1/2,t^{\prime})\,dt^{\prime}\,,\\
    \label{tvf}
\Theta(x,y;\tau,t_0) &=\frac{1}{\tau}\int^{t_0+\tau/2}_{t_0-\tau/2} \theta(x,y,z=1/2,t^{\prime})\, dt^{\prime}\,.
\end{align}  
Snapshot data is output periodically and $t_0$ is the time scale for this output interval (see the Supplementary Material for more details).
Both fields are transformed onto a polar wavevector grid in Fourier space giving $\hat{U}(k,k_{\phi};\tau, t_0)$ and 
$\hat{\Theta}(k,k_\phi;\tau, t_0)$. Azimuthally averaged Fourier spectra (see Fig. \ref{fig3}(b)) are given by
\begin{equation}
E_{\omega}(k;\tau, t_0)=\frac{1}{2\pi} \int_0^{2\pi} |\hat\omega(k,k_\phi;\tau, t_0)|^2 \,dk_\phi\,, 
\end{equation}
with $\hat\omega=\{\hat U,\hat \Theta\}$. All spectra $E_{\omega}(k;\tau, t_0)$ show a global maximum. An additional average over all
$t_0$ yields a unique maximum wavenumber $k^\ast_{U,\Theta}=2\pi/\hat\lambda_{U,\Theta}$ which depends on $Ra$ and $Pr$ as 
shown in Figs. \ref{fig3} (c,d). The wavelength $\hat\lambda_{U,\Theta}(Ra,Pr)/2$ is the characteristic mean width of the superstructure 
rolls as sketched in panel (a) of Fig. \ref{fig3}. We note that the spectra $E_{\omega}(k;\tau, t_0)$ do not vary significantly with $t_0$, 
in particular in respect to the maximum wavenumber $k^{\ast}$. The characteristic wavelengths in Figs. \ref{fig3}(c, d) are larger than 
the critical wavelength $\lambda_c=2\pi/k_c \approx 2$ at the onset of convection with $Ra_c=1708$ \cite{Chandrasekhar1961}. It is seen 
that the wavelength grows with $Ra$ at fixed $Pr$. The dependence on the Prandtl at fixed Rayleigh number in our data indicates a growth 
up to $Pr\sim 10$ and a subsequent decrease for even higher values which is in agreement with \cite{Hartlep2003} for smaller $\Gamma$. In the
Supplementary Material we demonstrate that nearly the same scales can be obtained by an analysis of the two-point correlation functions 
in physical space. 
  
\begin{figure*}
\centering
\includegraphics[width=0.75\textwidth]{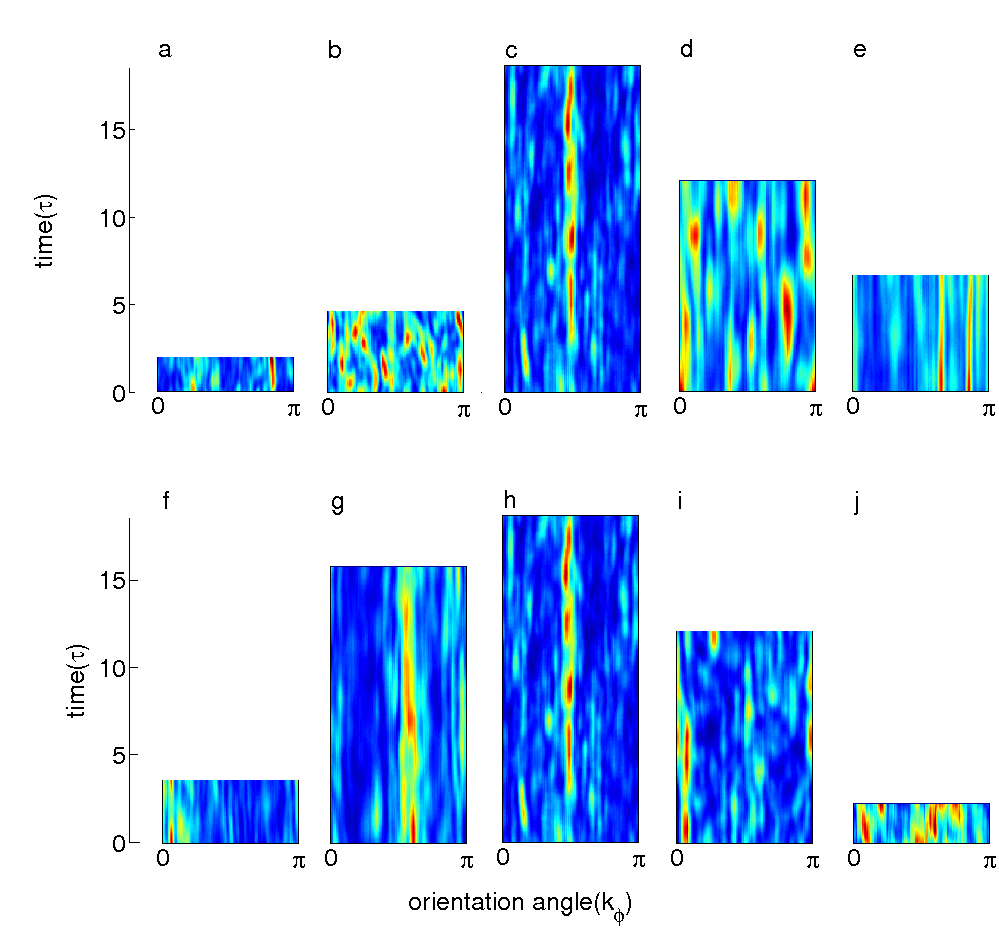}
\caption{{\bf Slow time evolution of superstructures.} Radially averaged temperature power spectrum $E_{\Theta}(k_{\phi}; \tau, t_0)$ versus $\tau$. We plot 
the angle in wavenumber space, $k_{\phi}=\arctan(k_y/k_x)$ between 0 and $\pi$. All panels are rescaled with their corresponding characteristic time
$\tau$ in correspondence with the table in the Supplementary Material. The top row shows data for $Ra=10^5$. (a) $Pr=0.005$, (b) $Pr=0.021$, (c) $Pr=0.7$, 
(d) $Pr=7$, and (e) $Pr=70$. The bottom row shows data for $Pr=0.7$. (f) $Ra=5\times 10^3$, (g) $Ra=10^4$, (h) $Ra=10^5$, (i) $Ra=10^6$, and (j) $Ra=10^7$.
Color scheme: deep blue corresponds to zero and red to the maximum value of each power spectrum.}
\label{fig4}
\end{figure*}

Interestingly, Fig. \ref{fig3} also shows that $\lambda_{\Theta} \gtrsim \lambda_{U}$. At the onset of convection, both wavelengths are exactly the 
same since both fields are perfectly synchronized in the midplane. Hot fluid is advected upwards ($\theta, u_z>0$) while cold fluid is brought downwards 
($\theta,u_z<0$). This perfect synchronicity breaks down with increasing $Ra$ since the temperature field is not only advected by vertical 
velocity component across the midplane, but also by rising horizontal velocity fluctuations. They expand the temperature patterns compared to those
of the vertical velocity component which manifests in a somewhat larger wavelength $\lambda_{\Theta}$. We quantified this effect by the calculation of a 
horizontal P\'{e}clet number $Pe_h=v_h H/\kappa$ based on a horizontal root mean square velocity in the midplane, $v_h=(\langle u_x^2\rangle+\langle u_y^2\rangle)^{1/2}$. The P\'{e}clet number is always  larger than 10 which underlines a dominance of convection in comparison to diffusion. 

With the characteristic width of the superstructure rolls (or cells) of $\hat{\lambda}_U/2$ determined, we can now define the characteristic turnover 
time for a fluid parcel. We estimate this time scale by an elliptical circumference, $\ell\approx \pi(a+b)$ with $a$ and $b$  (see again Fig. \ref{fig3}(a)) 
being the half-axes, and root mean square velocity of the turbulent flow. The characteristic time scale 
of the turbulent superstructures, beyond which the gradual evolution of the large-scale patterns proceeds, is given by
\begin{equation}
\tau(Ra,Pr)\approx 3\frac{\ell}{u_{rms}} \approx 3\frac{\pi \left(\frac{1}{4}\lambda_U+\frac{1}{2}H\right) } {\langle u_x^2+u_y^2+u_z^2\rangle_{V,t}^{1/2}} \,.
\label{tau}
\end{equation}
Figures~\ref{fig3}(e, f) display these computed times as a function of $Ra$ and $Pr$. The prefactor of 3 in Eq. (\ref{tau}) accounts for the 
fact that an individual fluid parcel is not perfectly circulating around in such a roll when the flow is turbulent. We tested that different prefactors 
of same order of magnitude do not change the results qualitatively (see also Supplementary Material). The characteristic time $\tau$ is found 
to be nearly unchanged at the fixed Prandtl number. It increases with $Pr$ at fixed $Ra$, remaining however always well below the upper 
bound, the dissipation time scale $T_d$ (see the table in the Supplementary Material). 
 
\vspace{0.5cm} 
\noindent
{\bf Radially averaged power spectra for slow superstructure evolution.}   
On time scales larger than $\tau$ the turbulent superstructure patterns are found to evolve by slow changes in orientation and topology. 
This can be quantified by an angular spectral analysis \cite{Zhong1992}. We take the radially averaged power spectrum of temperature 
$\Theta$ which is given by 
\begin{equation}
E_{\Theta}(k_\phi;\tau,t_0)=\frac{1}{k_m} \int_0^{k_m} |\hat\Theta(k,k_\phi;\tau,t_0)|^2\,dk\,, 
\label{radpower}
\end{equation}
and plot the spectra in Fig.~\ref{fig4} versus time $\tau$. The wavenumber $k_m$ in Eq.~(\ref{radpower}) denotes a cutoff with $k_m \gg k^{\ast}_{\Theta}$. 
Local maxima in this spectrum indicate now a preferential orientation of parallel rolls. The slow evolution of the turbulent superstructures 
becomes visible by the slow variation of the local maxima in the spectrum in all presented runs. We can identify in all cases a small number of local maxima 
that grow and then decay with time. As the old maxima decay, new ones set in that are shifted by discrete angles from the old ones. This suggests secondary modulations of the dominant roll pattern. We also see that for the highest $Pr$ the maxima persist for a very long period while they switch more rapidly 
in case of the lower $Pr$. This behaviour is reminiscent of cross-roll or skewed varicose instabilities that have been studied in detail in 
weakly nonlinear convection above onset \cite{Bodenschatz2000}. 

\begin{figure*}
\centering
\includegraphics[width=0.97\textwidth]{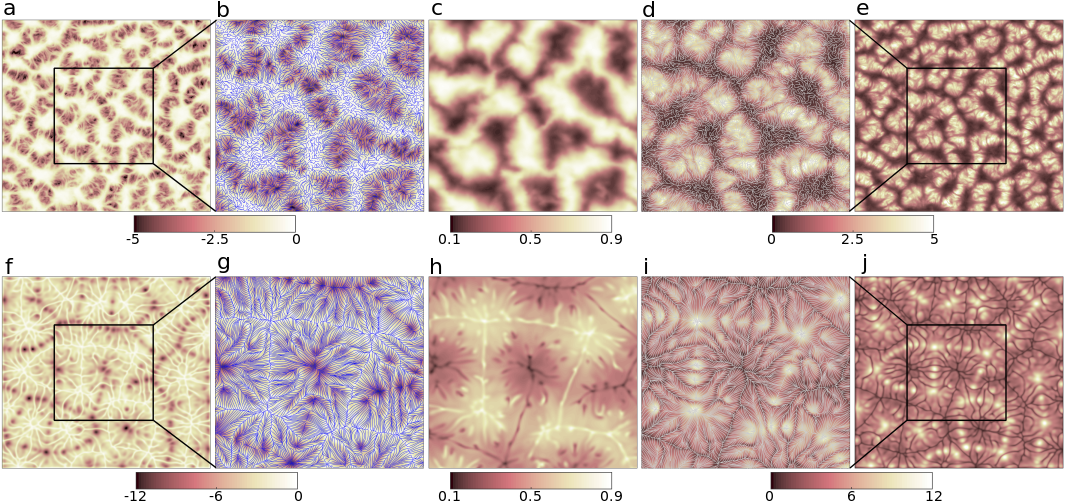}
\caption{{\bf Temperature field structure in boundary layer and midplane.} The data in top row (a -- e) of the figure are obtained for the case of 
$Pr=0.005$, the data in bottom row (f -- j) for $Pr=7$, both at $Ra=10^5$. Instantaneous snapshots are shown. (a, f) Contours of 
$\partial T/\partial z$ at $z=0$. (e, j) Contours of $-\partial T/\partial z$ at $z=1$. Boxes in these plots indicate the magnification region which 
is a quarter of the full cross section. Panels (b, d) and (g, i) are magnifications to (a,e) and (f, j), respectively. In addition, we display field lines 
of the corresponding skin friction field ${\bm s}$ in blue (white) for the bottom (top) plate. Panels (c, h) show the corresponding instantaneous 
temperature field $T$ in the midplane at $z=1/2$. Color bars are added to the corresponding figures. The size of panels (b--d) and (g--i) is the 
same.}
\label{fig5}
\end{figure*}

\vspace{0.5cm} 
\noindent
{\bf Connection of superstructures to boundary layers.}   
Figures \ref{fig3}(c, d) show that the characteristic scale of the superstructures varies with $Pr$ and $Ra$. For example, the growth of the 
wavelength with $Ra$ can be attributed to the increasingly erratic variations of the temperature filaments which in turn cause an effective increase 
of the size of the time-averaged structures in the Rayleigh number range that is monitored here. The trend with Prandtl number at fixed Rayleigh 
number is less obvious. Therefore Fig. \ref{fig5} compares the instantaneous temperature field structure in the boundaries at the top and bottom plates
with that in the midplane at $z=1/2$ for two different Prandtl numbers. We display the vertical temperature derivative $\partial T/\partial z$ at 
$z=0, 1$ in panels (a, e, f, j) of Fig. \ref{fig5}. This field is one way to highlight the thermal plume ridges \cite{Shishkina2005}. As expected, 
thin filaments and subfilaments are observed for a higher $Pr$ while the derivative contours appear somewhat blurred and coarse grained for 
the lowest Prandtl number. 

Panels (b, d, g, i) of Fig. \ref{fig5} display a zoom of the same data together with the field lines of the skin friction field ${\bm s}=(\partial_z u_x, 
\partial_z u_y)$ at the plates. This two-dimensional vector field is composed of the two non-vanishing components of the velocity 
gradient tensor at $z=0,1$. It contains sources and sinks and is fully determined by its critical points, ${\bm s}=0$ \cite{Chong2012,Bandaru2015}. 
These critical points are either unstable nodes, stable nodes or saddles, and much less frequently unstable and stable foci. Groups of saddles and 
stable nodes are correlated with local regions of the formation of dominant plumes while unstable nodes are mostly found where colder (hotter) fluid 
impacts the bottom (top) plate. This is very clearly visible for $Pr=7$ in the bottom row of the figure, but does also hold for the  
low-Prandtl-number data displayed here. The structures at the top plate display the same plume ridges, but are shifted by a roll-length when compared 
with those at the bottom plate, as is expected for a system of parallel rolls (see also Supplementary Material). 

\begin{figure}
\centering
\includegraphics[width=0.45\textwidth]{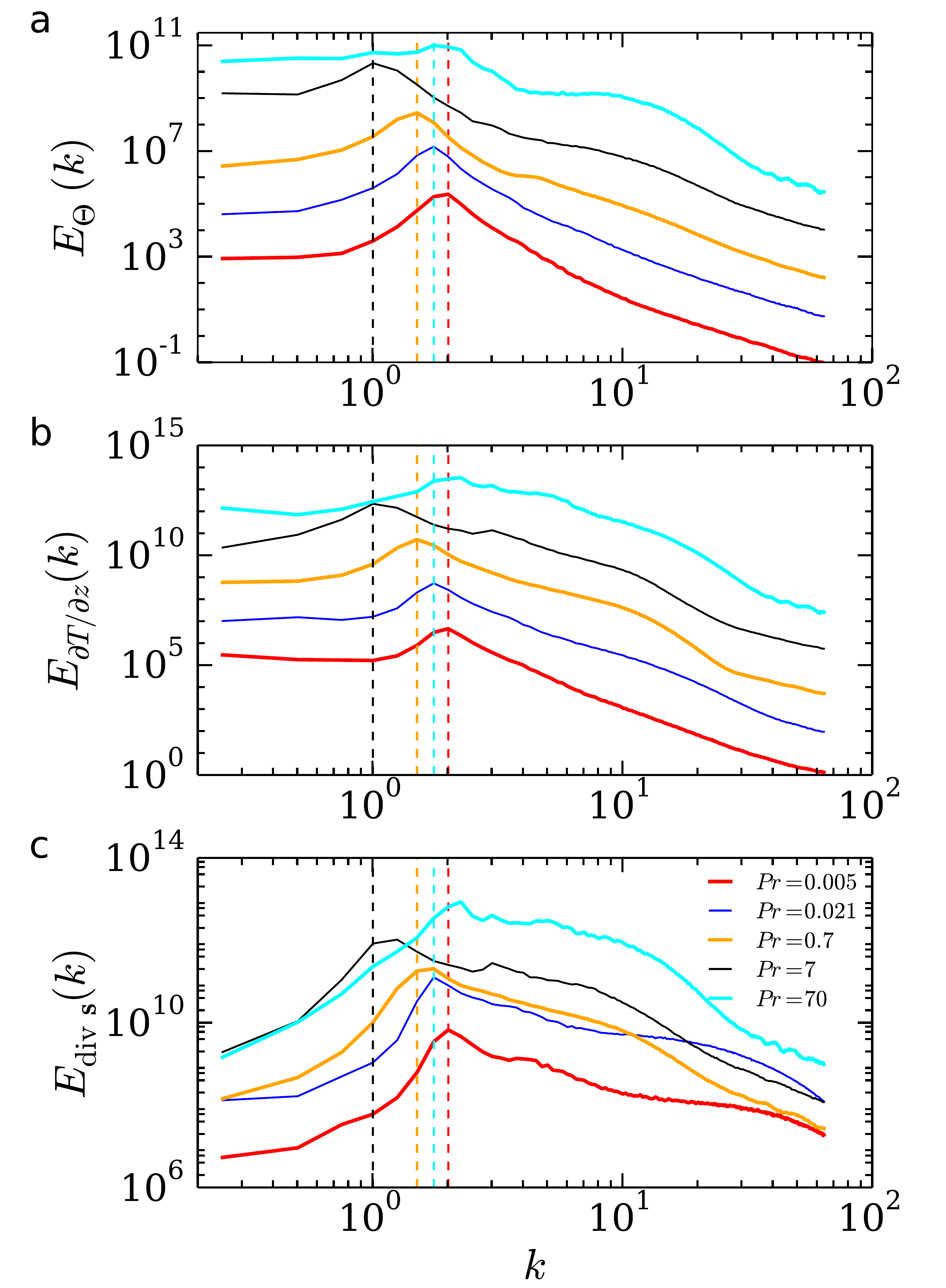}
\caption{{\bf Scale correlations of bulk and boundary layer.} Spectra of temperature $\Theta$ in midplane (a), of vertical temperature derivative 
$\partial_z T$ at bottom and top plates (b), and of skin friction field divergence at bottom and top plates (c) are compared for five different Prandtl numbers
as indicated by the legend. The vertical dashed lines denote the maximum wavenumbers $k^{\ast}_{\Theta}$ and are replotted in panels (b) and (c). 
The spectra are averaged over all snapshots. The dashed lines for $Pr=0.021$ and 70 (blue and cyan) collapse.}
\label{fig6}
\end{figure}

The panels (c, h) of Fig. \ref{fig5} show the instantaneous temperature field $T$ in
the midplane with local maxima and minima exactly where the hot and
cold plume ridges are present at the plates, respectively. These
dominant ridges are the ones that persist as the superstructures once
the time-averaging over $\tau$ is performed. Figure \ref{fig6}
demonstrates also that the turbulent superstructures are directly
connected to the strongest thermal plumes in the boundary layers. This
plume formation process is determined by two aspects: (i) the
molecular diffusivity of the temperature field (and the resulting
differences in the thicknesses of thermal and viscous boundary layers)
and (ii) the typical variation scale of the horizontal velocity field
near the walls that forms the plume ridges by temperature field
advection. While the first aspect will affect the shape of the plume
ridges and thus the characteristic thickness scale of the
local temperature maxima and minima in the midplane, the second one is
directly connected to the spacing of the dominant temperature
structures in the midplane and thus the width of the large-scale
circulation rolls and cells that fill the layer.  The divergence of
the skin friction field which is given at the bottom plate by
\begin{equation}
\mbox{div} {\bm s}=\frac{\partial^2 u_x}{\partial x\,\partial z}\Bigg|_{z=0}+ \frac{\partial^2 u_y}{\partial y\,\partial z}\Bigg|_{z=0}
=-\frac{\partial u_z^2}{\partial z^2}\Bigg|_{z=0} \,,  
\label{divs}
\end{equation}
can be considered as a blueprint of the alternating impact (source with $\mbox{div} {\bm s}>0$) and ridge formation (sink with $\mbox{div} {\bm s}< 0$) 
regions. The skin friction field is thus a key to understanding the clustering of thermal plumes near the wall, a phenomenon which has been reported for example 
in \cite{Parodi2004}. The same picture holds at the top plate.          
   
Figure \ref{fig6} underlines this correlation by means of the power spectra of the temperature in the midplane, the vertical temperature derivative 
at the plates and the divergence of the skin friction field. We have applied again the sliding time average over $\tau$. All three spectra are found 
to peak at the same scale (except $Pr\ge 7$ where the scales however are still comparable). Our result is thus robust with respect to $Pr$ and 
underlines that the same dynamical processes are at work for all Prandtl numbers.  As seen in Fig. \ref{fig6}, the characteristic scale of the skin 
friction divergence is expected to decrease when the Prandtl number gets smaller. It is documented in refs. \cite{Schumacher2016,Scheel2016} 
that the Reynolds number increases significantly when $Pr$ decreases at constant $Ra$ thus indicating a much more vigorous fluid turbulence, 
both in the bulk and in the boundary layers (see also Supplementary Material). Thus the spatially extended advection patches of the horizontal 
velocity field, as visible in the magnification for the case for $Pr=7$ in Fig. \ref{fig5} (g, i), will not persist for low-Prandtl-number convection.

\section*{Discussion}     
Our main motivation was to study the large-scale patterns in turbulent convection which are termed turbulent superstructures. We then analysed the 
characteristic length and time scales associated with these turbulent superstructures as a function of Rayleigh and Prandtl numbers and found 
a separation between large-scale, slowly evolving structures and  small-scale, rapidly turning vortices and filaments. The system that we have chosen is the 
simplest setting for a turbulent flow that is initiated by temperature differences, a Rayleigh-B\'{e}nard convection flow between uniformly heated 
and cooled plates. This flow has already been studied intensively with respect to pattern formation in the weakly nonlinear regime above the onset of 
convection at $Ra=Ra_c$, as documented in the cited reviews \cite{Busse1978,Cross1993,Bodenschatz2000}. Our study shows that patterns of 
rolls and cells continue to exist into the fully turbulent and time-dependent flow regime once the small-scale fluctuations of the temperature and 
velocity fields are removed. 

Prandtl numbers that vary here over more than four orders of magnitude change the character of convective turbulence drastically from a 
highly inertia-dominated Kolmogorov-type turbulence at the lowest $Pr$ to a fine-structured convection at the highest $Pr$. This results in a strong 
dependence of the characteristic spatial and temporal separation scales that are necessary to describe the gradual large-scale evolution of the flow at
hand. These spatial separation scales are found to continuously increase up to $Pr\lesssim 10$ and to decay for $Pr\gtrsim 10$ for the parameter values 
that we were able to cover here which is in agreement with \cite{Hartlep2003}. A saturation of the characteristic scale might occur for the opposite 
limit, $Pr\to 0$. Our data indicate such a behavior which is supported by previous studies at zero-Prandtl convection by Thual \cite{Thual1992}. There
they found only small differences between $Pr=0.025$ and the singular limit $Pr=0$. However these former studies have been conducted in much 
smaller boxes at significantly smaller spectral resolutions.  

A further interesting observation that was made in the present study is the connection between the mean scales of the turbulent 
superstructure patterns analysed in the midplane and those of the near-wall flows. Our analysis suggests that the characteristic scales 
of large-scale superstructures are correlated with the thermal plume ridges in the boundary layers. We showed for all $Pr$ that 
the maximum wavenumber of the temperature spectrum in the midplane $k^{\ast}_{\Theta}$ nearly perfectly coincides with the wavenumber at which the power 
spectrum of the divergence of the skin friction field peaks. The latter wavenumber characterizes the mean distance of impact (div ${\bm s} >0$)
and ejection (div ${\bm s} <0$) regions at the walls. It is thus the characteristic variation scale of the horizontal velocity field that advects the hot (cold) fluid 
together at the bottom (top) boundary to form prominent thermal plume ridges. The interplay between the thermal and viscous boundary layers of different 
thicknesses could thus be responsible for the variation of the characteristic superstructure scale with growing $Pr$. The viscous boundary layer becomes 
ever thicker as $Pr$ increases and velocity fluctuations decrease thus generating more coherent advection patterns. Competing boundary layers 
that control transport and structure formation in convection flows have been discussed in other settings, for example in ref. \cite{King2009} 
for rapidly rotating convection.
 
The characteristic superstructure scales which we have detected in the present work suggest a scale separation for convective turbulence.
There is the fast convective motion below the characteristic width of individual circulation rolls or cells on times smaller than several tens of 
 free-fall times. Then  after the removal of the small-scale turbulence, the large-scale patterns of rolls are revealed and these fill the 
whole layer and vary slowly on time scales larger than a few hundreds of free-fall times. The latter dynamic processes can be of interest for a global effective
description of mesoscale convection phenomena in atmospheric turbulence \cite{Randall2003} or of pattern formation in a scale range between solar granulation and 
supergranulation \cite{Rincon2017}.  In contrast to rapidly rotating convection flows or magnetoconvection in the presence of strong  external magnetic fields, 
the present RBC flow permits a mathematically rigorous asymptotic expansion that generates simplified equations for the dynamics of these patterns 
(see e.g. \cite{Stellmach2014}). The unresolved dynamics at the fine and fast scales below $\lambda_{\Theta,U}$ and $\tau$ will  be modeled
empirically. This is being further investigated and will be reported elsewhere.      
    
\section*{Methods}

\noindent
{\bf Boussinesq equations and numerical method.}
We solve the coupled three-dimensional equations of motion for velocity field $u_i$ and temperature field $T$ in the Boussinesq approximation
of thermal convection: 
\begin{align}
\label{ceq}
\frac{\partial u_i}{\partial x_i}&=0\,,\\
\label{nseq}
\frac{\partial  u_i}{\partial  t}+u_j \frac{\partial u_i}{\partial x_j}
&=-\frac{\partial p}{\partial x_i}+\sqrt{\frac{Pr}{Ra}} \frac{\partial^2 u_i}{\partial x_j^2}+  T \delta_{i3}\,,\\
\frac{\partial  T}{\partial  t}+u_j \frac{\partial T}{\partial x_j}
&=\frac{1}{\sqrt{Ra Pr}} \frac{\partial^2 T}{\partial x_j^2}\,,
\label{pseq}
\end{align}
with Rayleigh number $Ra=g\alpha\Delta T H^3/(\nu\kappa)$ and Prandtl number $Pr=\nu/\kappa$.
The equations are made dimensionless by cell height $H$, free-fall velocity $U_f=\sqrt{g \alpha \Delta T H}$ and the imposed 
temperature difference $\Delta T$ between bottom and top plates. The aspect ratio $\Gamma=L/H=25$ with the cell length $L$. 
The variable $g$ stands for the  acceleration due to gravity, $\alpha$ is the thermal expansion coefficient, $\nu$ is the kinematic viscosity, and 
$\kappa$ is the thermal diffusivity. No-slip boundary conditions for the fluid are applied at all walls. The sidewalls are thermally insulated and the 
top and bottom plates are held at constant dimensionless temperatures $T=0$ and 1, respectively. The equations are numerically solved by 
the Nek5000 spectral element method package \cite{nek5000}. We have two series of direct numerical simulations:
six runs at $Pr=0.005$, 0.0021, 0.7, 7, 35, and 70 for $Ra=10^5$ and six runs at $Pr=0.7$ for $Ra=2.3\times 10^3, 5\times 10^3, 10^4, 10^5, 10^6$ and $10^7$.
Details on the size of the simulations as well as some characteristic parameters of the simulations can be found in a comprehensive table in
the Supplementary Material.

\section*{Acknowledgements.} 
AP and JDS acknowledge support by the Deutsche For\-schungs\-gemeinschaft within the Priority Programme
Turbulent Superstructures under Grant No. SPP 1881. We acknowledge supercomputing time at the Blue  Gene/Q JUQUEEN at the J\"ulich 
Supercomputing Centre by large-scale project HIL12 of the John von Neumann Institute for Computing and at the SuperMUC Cluster at
the Leibniz Supercomputing Centre Garching by large-scale project pr62se. 

\section*{Author contributions}
All three authors made significant contributions to this work. All authors designed the numerical experiments and analysed the data. 
AP and JS ran the production simulations at the supercomputing sites in Garching and J\"ulich. All authors discussed the results and 
wrote the paper together.

\end{document}


\newcommand{\joerg}[1]{\textcolor{red}{#1}}
\newcommand{\janet}[1]{\textcolor{blue}{#1}}

\title{Supplementary Material:\\
Turbulent superstructures in Rayleigh-B\'{e}nard convection}
\author{Ambrish Pandey}
\affiliation{Institut f\"ur Thermo- und Fluiddynamik, Technische Universit\"at Ilmenau, Postfach 100565, D-98684 Ilmenau, Germany}
\author{Janet D. Scheel}
\affiliation{Department of Physics, Occidental College, 1600 Campus Road, M21, Los Angeles, California 90041, USA}
\author{J\"org Schumacher}
\affiliation{Institut f\"ur Thermo- und Fluiddynamik, Technische Universit\"at Ilmenau, Postfach 100565, D-98684 Ilmenau, Germany}
\date{\today}
\maketitle

\section{Simulations details and further data analysis}
\subsection{Parameters of the simulation runs}
Here, we provide further details on the numerical simulations with the Nek5000 spectral element package \cite{nek5000}
and on the data analysis. In Table 1, we summarize the most important parameters of all simulations. The turbulent momentum 
and heat transfer are quantified by the Reynolds and Nusselt numbers, $Re$ and $Nu$, respectively. They are defined in 
dimensionless form as 
\begin{equation}
Re=\langle u_x^2+u_y^2+u_z^2\rangle^{1/2}_{V,t}\,\sqrt{\frac{Ra}{Pr}}\,,\quad\quad\mbox{and}\quad\quad Nu=1+\sqrt{Ra Pr}\langle u_z T\rangle_{V,t}\,,
\end{equation}
with $\langle\cdot\rangle_{V,t}$ being a full volume-time average. We have verified that the spectral resolution is sufficient
in correspondence with criteria that are summarized in \cite{Scheel2013}. We compared two pairs of runs at different spectral 
resolution and tested that the results for the global transport of heat and momentum across the layer are the same. These are runs 
3 and 3a at $Pr=0.7$ and $Ra=10^5$ as well as runs 4 and 4a at $Pr=7$ and $Ra=10^5$ (see Table I of Supplementary Material). 
We also verified that the vertical profiles of the plane- and time-averaged kinetic energy and thermal dissipation rates exhibit smooth curves across 
the element boundaries \cite{Scheel2013}.

\begin{figure*}
\centering
\includegraphics[width=0.98\textwidth]{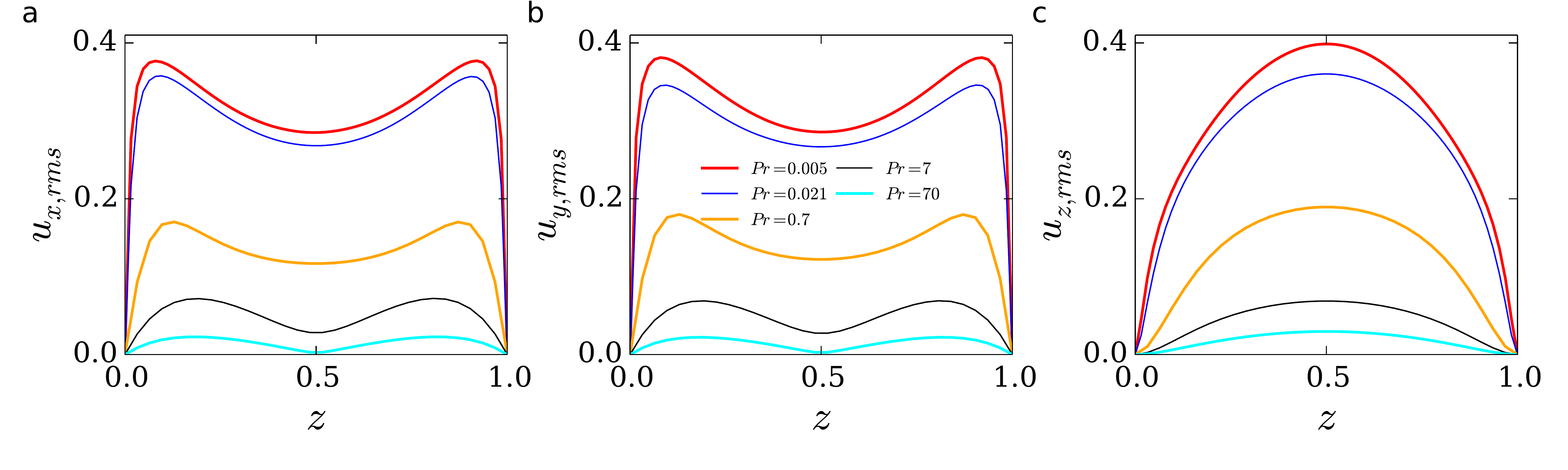}
\caption{Vertical profiles of the root mean square velocity fluctuations. The data are obtained by a combined temporal and plane average
over the whole cross section area. (a) $u_{x,rms}(z)$, (b) $u_{y,rms}(z)$, and (c) $u_{z,rms}(z)$. The corresponding Prandtl numbers are 
indicated in the figure. All data are collected at $Ra=10^5$.}
\label{fig1s}
\end{figure*}

Figure \ref{fig1s} shows the profiles of 
the root mean square fluctuations of the three velocity components. They are given by 
\begin{align}
u_{x,rms}(z)&=\langle u_x^2\rangle^{1/2}_{A,t}\,,\\
u_{y,rms}(z)&=\langle u_y^2\rangle^{1/2}_{A,t}\,,\\
u_{z,rms}(z)&=\langle u_z^2\rangle^{1/2}_{A,t}\,,
\end{align}
where $\langle\cdot\rangle_{A,t}$ denotes an average over horizontal $x-y$ planes at fixed $z$ and time. 
It can be seen that the profiles for $u_{x,rms}$ and $u_{y,rms}$ agree well, 
as expected in a horizontally extended domain. The convection flow at $Ra=10^5$ is in a turbulent state for Prandtl numbers $Pr\le 0.7$.
For $Pr=7$ the fields are in a chaotically time-dependent state while for $Pr=70$ temperature and velocity fields vary very slowly 
with respect to time. The magnitude of all three profiles $u_{i,rms}$ is significantly reduced for $Pr=70$ as seen in Fig. \ref{fig1s}.   
\begin{table*}
\begin{ruledtabular}
\begin{center}
\begin{tabular}{rccccccccccccc}
    &  $Pr$  &  $Ra$  &   $N_e$  &   $N$  &  $Nu$  &   $Re$  &  $u_{rms}$ & $N_s$ &  $\tau_{\text{total}}$ & $\tau$ & $t_{\kappa}$ & $t_{\nu}$\\
\hline
1  &  0.005 &  $10^5$ &   2,367,488  & 11  & $1.9 \pm 0.01$  & $2491 \pm 20$  &  $0.56 \pm 0.004$ &  63  &  60  & 21 & 22 & 4472\\                      
2  &  0.021 &  $10^5$ &   2,367,488  &  7   & $2.6 \pm 0.01$ & $1120 \pm 8$  & $0.51 \pm 0.004$ &  69   &  156 & 27 & 46 & 2182 \\
3  &  0.7     &  $10^5$ &   2,367,488  &  3   & $4.2 \pm 0.02$  &  $92 \pm 0.4$  &  $0.24 \pm 0.001$ & 117  &  1159 & 57 &265 & 378\\
3a  &  0.7     &  $10^5$ &   1,352,000  &  5   & $4.3 \pm 0.02$  &  $92 \pm 0.4$  &   $0.24 \pm 0.001$  &  278  & 1108  & 57  & 265 &  378\\
4  &  7        &  $10^5$  &  2,367,488  &  3   & $4.1 \pm 0.01$  &  $11 \pm 0.04$ &  $0.09 \pm 0.001$  & 150  &  2979 & 192 & 837 & 120\\
4a  &  7        &  $10^5$  &  1,352,000  &  5   & $4.1 \pm 0.01$  &  $11 \pm 0.03$ &  $0.09 \pm 0.001$  & 268  &  2670 & 207 & 837 & 120\\
5  &  35      &  $10^5$  &  1,352,000  &  5  & $4.5 \pm 0.01$ & $2.3 \pm 0.005$ & $0.04 \pm 0.0001$ & 343 & 3420 & 300 & 1871 & 53 \\
6  &  70      &  $10^5$  &  1,352,000  &  5   & $4.6 \pm 0.01$ & $1.2 \pm 0.002$ & $0.03 \pm 0.0001$ & 337 & 3360 & 375 & 2646 & 38 \\
     &    & & & & & & & & \\
7  &  0.7     &  $2.3\times 10^3$  &  2,367,488  &  3   & $1.3 \pm 0.001$  & $6.1 \pm 0.04$ & $0.11 \pm 0.001$  & 74 & 1460 & 93 & 40 & 57 \\
8  &  0.7     &  $5.0\times 10^3$  &  2,367,488  &  3   &  $1.8 \pm 0.005$ & $14 \pm 0.2$ & $0.17 \pm 0.002$ &  51 & 250 & 54 & 59 &	85\\
9  &  0.7  &  $10^4$  &  1,352,000  &  5 & $2.2 \pm 0.01$ & $24 \pm 0.2$ & $0.20 \pm 0.002$ & 240 & 1195 & 72 & 84 & 120\\
10  &  0.7     &  $10^6$  &  2,367,488  &  7   &  $8.1 \pm 0.03$ & $290 \pm 1$ & $0.24 \pm 0.001$ & 392 & 1292 & 66 & 837 & 1195\\
11 &   0.7     &  $10^7$  &  2,367,488  & 11  & $15.5 \pm 0.06$ & $864 \pm 3$ & $0.23 \pm 0.001$ & 127 & 248 & 72 & 2646 & 3780\\
\end{tabular}  
\caption{Parameters of the 11 different spectral element simulations. We display the Prandtl number $Pr$, the 
Rayleigh number $Ra$, the number of spectral elements $N_e$, the polynomial order of the expansion in each 
of the three space directions $N$ on each element, the Nusselt number $Nu$,  the Reynolds number $Re$, the 
root mean square velocity $u_{rms}=\langle u_i^2\rangle^{1/2}_{V,t}$, the number of statistically independent snapshots 
$N_s$, and the total runtime $\tau_{\text{total}}$ in units of the free-fall time $T_f=H/U_f$. We also list the turnover time $\tau$, 
the vertical diffusion time $t_{\kappa}=H^2/\kappa=\sqrt{Ra Pr}\,T_f$,  and the vertical viscous time $t_{\nu}=H^2/\nu=\sqrt{Ra/Pr}\,T_f$. 
The snapshot separation is found from $\tau_{\text total}/(N_s-1)$ and $t_0$ is given in  these units.
Runs 3a and 4a are conducted at different spectral resolution and compared with runs 3 and 4, respectively.}
\end{center}
\end{ruledtabular}
\end{table*}
\begin{figure*}
\centering
\includegraphics[width=0.85\textwidth]{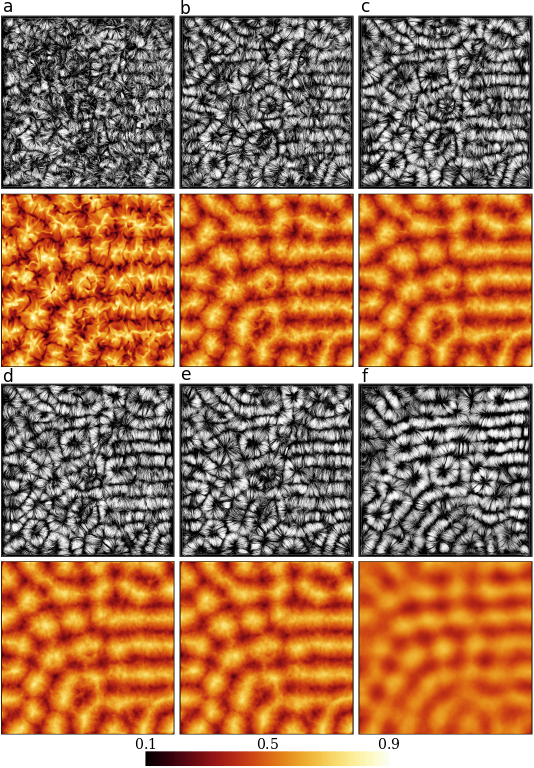}
\caption{Successively longer time averaging of the data for the simulation run at $Ra=10^5$ and $Pr=0.7$.
Velocity field (top) and temperature in the midplane (bottom) are displayed. (a) Instantaneous snapshot. (b--f) Time 
averaged fields for a successively longer interval length of $\tau/3$, $2\tau/3$, $\tau$, $4\tau/3$, and $19.4\tau$.
It can be seen that the patterns remain relatively independent of the interval length for panels (d) and (e) which
corresponds to averages over 57 and 76 $T_f$, respectively. Panel (f) displays the time average over the 
full time integration.}
\label{fig2s}
\end{figure*}
\begin{figure*}
\centering
\includegraphics[width = 0.85\textwidth]{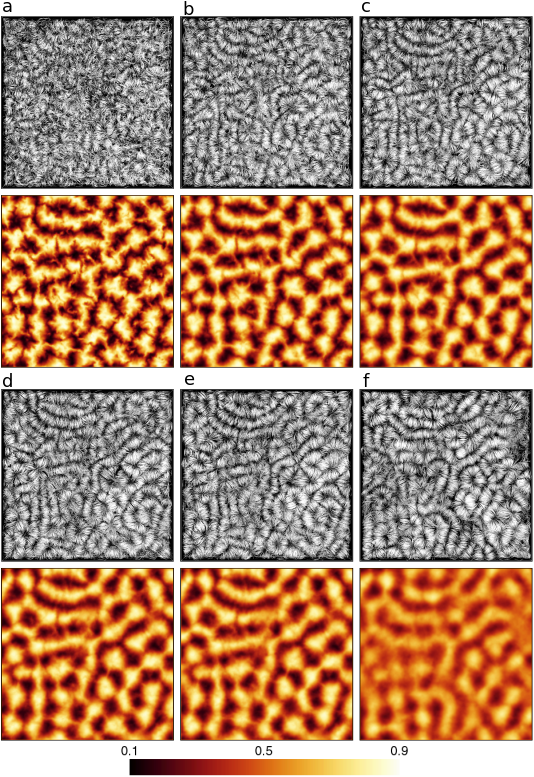}
\caption{Successively longer time averaging of the data for the simulation run at $Ra=10^5$ and $Pr=0.021$.
Velocity field (top) and temperature in the midplane (bottom) are displayed. (a) Instantaneous snapshot. (b--f) Time 
averaged fields for a successively longer interval length of $\tau/3$, $2\tau/3$, $\tau$, $4\tau/3$, and $5.5\tau$.
It can be seen that the patterns remain relatively independent of the interval length for panels (d) and (e) which 
corresponds to averages over 27 and 36 $T_f$, respectively. Panel (f) displays again the time average over the 
full time integration.}
\label{fig3s}
\end{figure*}

\subsection{Superstructures as a function of the interval length of time average}
As mentioned in the main text, the time averaging window should be long enough to remove small-scale 
turbulent fluctuations. The appropriate time scale is determined from the characteristic 
horizontal pattern scale and the amplitude of the turbulent velocity fluctuations (see Eq. (7) of the main 
text). Both of these quantities are only determined after the simulation is finished. For each parameter set, we tested 
how the patterns are affected by a successively longer time average interval.  Figures \ref{fig2s}
and \ref{fig3s} summarize these results for the runs at $Pr=0.7$ and 0.021, respectively. Panel (a) in both figures is
an instantaneous snapshot, the averaging time interval is successively increased and centered around the original 
snapshot. Compared to panels (a) in both figures it can be seen that the patterns in panels (d) and (e) remain relatively robust. While the 
large-scale structures of the temperature are visible for shorter time intervals already, they become clearly observable for the 
velocity only after a certain length ($\sim \tau$) of the interval. For even longer time averages in turn, the temperature structures begin to become  
washed out while velocity patterns do not disappear, but change gradually.     
\begin{figure*}
\centering
\includegraphics[width=0.9\textwidth]{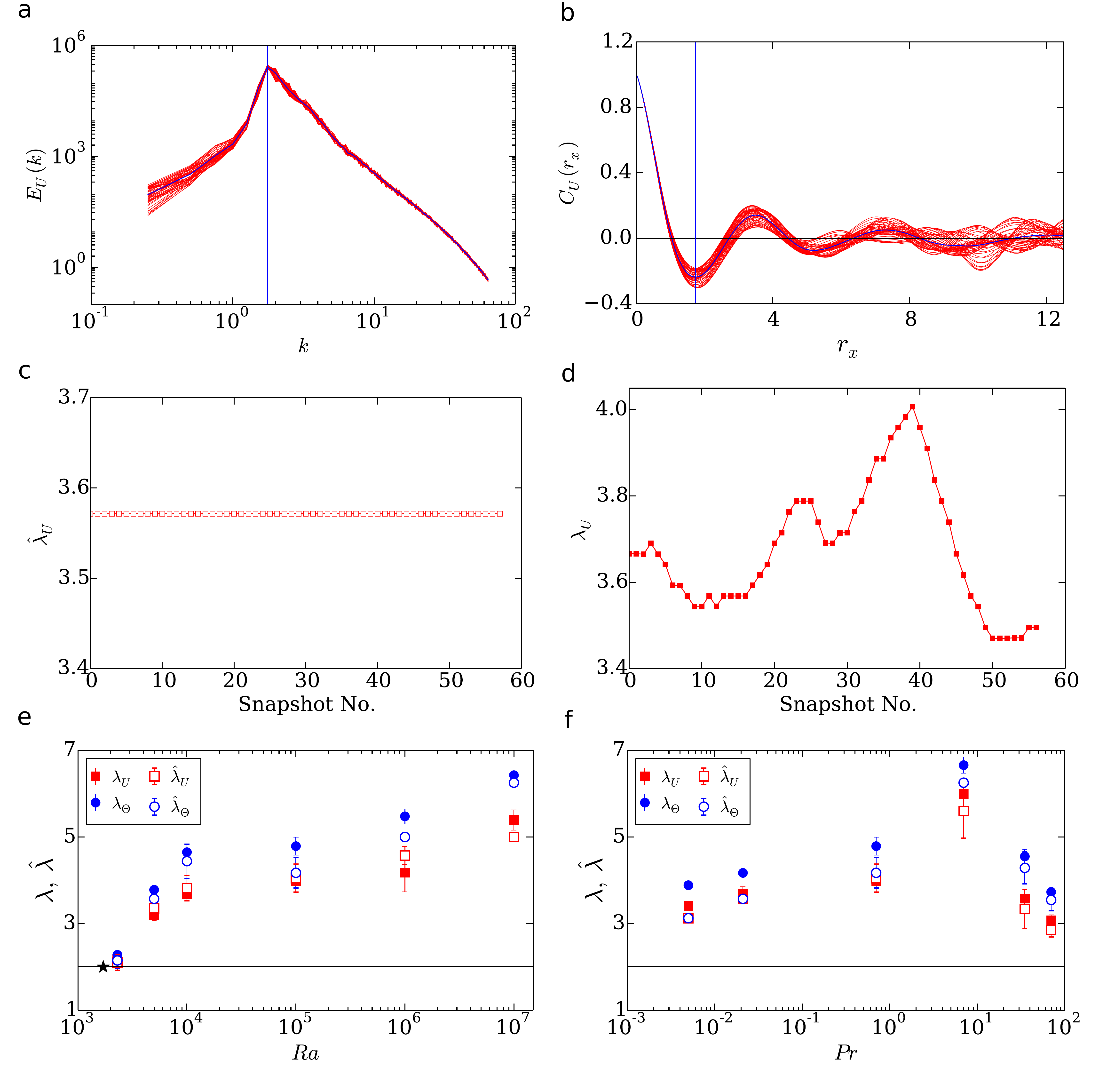}
\caption{Complementary scale analysis in physical space by means of correlation functions. Panels (a--d) show the data for one example 
run at $Pr=0.021$ and $Ra=10^5$. (a) Azimuthally averaged spectra of $\hat{U}(k_x, k_y; \tau, t_0)$ (red curves) and the mean over different $t_0$ of all these 
spectra (blue curve). The vertical solid line indicates the wavenumber that corresponds to the maximum of the averaged spectrum. (b) Spatial correlation 
functions taken for $U(x, y; \tau, t_0)$ in physical space along $x$-direction. Again, correlation functions at different 
$t_0$ (red curves) and the mean of all correlations (blue curve) are shown. (c, d) Resulting characteristic scales of the superstructures, $\hat{\lambda}_U$ 
and $\lambda_U$ versus snapshot number, i.e., as a function of $t_0$. (e, f) Summary of characteristic scales for $U$ and $\Theta$ which are 
obtained in Fourier ($\hat \lambda$) and physical space ($\lambda$), respectively. The horizontal solid lines in panels (e) and (f) 
stand for the critical wavelength at the onset of convection. Star symbol in panel (e) stands for $Ra_c=1708$.}
\label{fig4s}
\end{figure*}
\begin{figure*}
\centering
\includegraphics[width=0.9\textwidth]{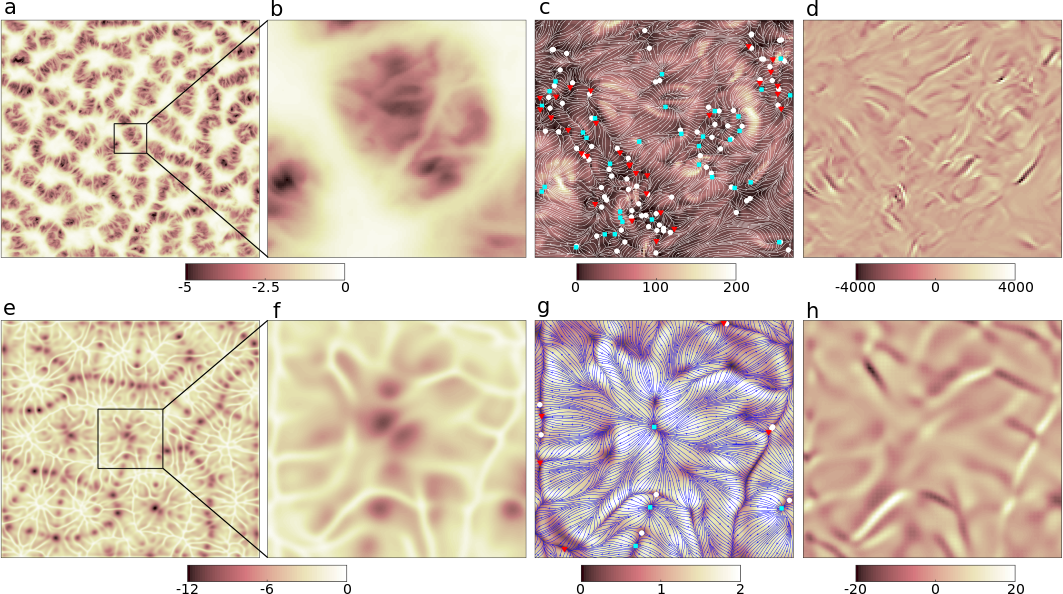}
\caption{Magnification of the structure of the skin friction field at the bottom plate. Panels (a--d) are for $Pr=0.005$ and (e--h) for $Pr=7$. 
Both data sets are obtained for $Ra=10^5$ and correspond to the data shown in Fig. 6 of the main text. Panels (a, e) show the derivative 
$\partial T/\partial z$ at the plate $z=0$. Panels (b, f) replot these data in the magnified section. For the top panel (b) $-1.6\le x,y\le 1.6$ and 
for the bottom panel (f) $-3.1\le x,y\le 3.1$ are taken. Panels (c, g) display the field lines of the 
skin friction field and the critical points as symbols. White circles correspond to saddles, cyan squares to unstable nodes or foci 
and red triangles to stable nodes or foci. The colored background stands for the magnitude $s=|{\bm s}|$. Panels (d, h) show contours of the 
divergence of the skin friction field. }
\label{fig5s}
\end{figure*}
\subsection{Complementary pattern scale analysis in physical space.}
Figure \ref{fig4s} provides results that were obtained by a complementary analysis of the spatial correlation functions in physical space.
Similar to the spectral analysis in the main text, we average the vertical velocity $u_z$ and temperature deviation $\theta$ over a time interval
$\tau$ and thus generate a sequence of fields $U(x, y; \tau, t_0)$ and $\Theta(x, y; \tau, t_0)$ at different times $t_0$ (see also Eqns. (4) and (5) 
of the main text). The correlation functions are given by (here along $x$ direction)
\begin{equation}
C_{\omega}(r; \tau, t_0)=\frac{\langle\omega(x+r,y;\tau,t_0)\omega(x,y;\tau,t_0)\rangle_{x,y}}{\langle \omega^2\rangle_{x,y}}\,,  
\end{equation}
with $\omega=\{U,\Theta\}$ taken at $z=1/2$. Correlations along the $y$ direction are defined similarly. A well-defined minimum of $C_{\omega}$ 
with strongest anti-correlation is found for all data sets at $r^{\ast}_{U,\Theta}=\lambda_{U,\Theta}/2$. Figure \ref{fig4s}(b) displays the obtained results
for $U$. For completeness, we replot the corresponding spectra in Fig. \ref{fig4s}(a) and indicate the maximum 
\begin{equation}
k^{\ast}_U=\frac{2\pi}{\hat{\lambda}_U}\,.
\end{equation} 
The analysis for $\Theta$ proceeds in exactly the same way. The analysis along $x$ and $y$ directions is combined. Figures \ref{fig4s}(c,d)
display the resulting characteristic scales obtained in Fourier space by means of the azimuthally averaged spectra and in physical space by 
means of correlations along $x$ and $y$ directions versus $t_0$ or starting snapshot number of time average window $\tau$. The characteristic
superstructure scales $\lambda_{U,\Theta}$ and $\hat{\lambda}_{U,\Theta}$ are found by averaging over all snapshots. The results are compared in Figs.
\ref{fig4s}(e, f). It is seen that the results for the characteristic scales which we obtained in two ways agree very well.
 
\subsection{Skin friction field and critical points}
Figure \ref{fig5s} adds further information on the structure of the skin friction field and is thus related to Fig. 6 of the main text. We display here a 
stronger zoom and add the spatial distribution of the critical points which are obtained in this small section of the bottom plate. The divergence 
of the skin friction field is also shown as a contour plot. The direct comparison of both data sets illustrates how the characteristic horizontal scales of the 
boundary layer dynamics of the velocity and temperature fields become smaller as the Prandtl number decreases from 7 to 0.005. Note that the 
magnification section in case of $Pr=0.005$ has to be chosen smaller in order to display all  features. It can also be seen that the number of critical 
points increases significantly as $Pr$ gets smaller.